\documentclass[twocolumn,showpacs,nofootinbib]{revtex4-1}
\usepackage{epsfig,amssymb,amsmath,latexsym}
\usepackage{pifont}
\usepackage{ulem}
\usepackage{hyperref}
\usepackage{subfigure}
\usepackage{multirow}

\usepackage{graphicx}
\usepackage{dcolumn}
\usepackage{bm}
\usepackage{epsfig}
\usepackage{color}

\usepackage{color}
\usepackage{hyperref}

\hypersetup{
    colorlinks=true,
    linkcolor=red,
    citecolor=blue,
}

\definecolor{nicegreen}{rgb}{0.1,0.5,0.1}

\def\fun#1#2{\lower3.6pt\vbox{\baselineskip0pt\lineskip.9pt
  \ialign{$\mathsurround=0pt#1\hfil##\hfil$\crcr#2\crcr\sim\crcr}}}
\def\simgt{\mathrel{\lower0.6ex\hbox{$\buildrel {\textstyle >}
 \over {\scriptstyle \sim}$}}}
\def\simlt{\mathrel{\lower0.6ex\hbox{$\buildrel {\textstyle <}
 \over {\scriptstyle \sim}$}}}

\def\bea{\begin{eqnarray}}
\def\eea{\end{eqnarray}}
\def\be{\begin{equation}}
\def\ee{\end{equation}}

\input epsf

\def\be{\begin{equation}}
\def\ee{\end{equation}}
\def\ba{\begin{eqnarray}}
\def\ea{\end{eqnarray}}

\newcommand{\no}{\noindent}

\begin{document}

\preprint{}

\title{Implications of the pulsar timing array detections for massive black hole mergers in the LISA band}

\author{Enrico Barausse$^{1,2}$, Kallol Dey$^{3}$, Marco Crisostomi$^{1,2}$, Akshay Panayada$^3$, Sylvain Marsat$^4$, Soumen Basak$^3$}
	
\affiliation{
$^1$SISSA, Via Bonomea 265, 34136 Trieste, Italy and INFN Sezione di Trieste \\
$^2$IFPU - Institute for Fundamental Physics of the Universe, Via Beirut 2, 34014 Trieste, Italy \\
$^3$School of Physics, Indian Institute of Science Education and Research Thiruvananthapuram,\\ Maruthamala PO, Vithura, Thiruvananthapuram 695551, Kerala, India\\
$^4$Laboratoire des 2 Infinis - Toulouse (L2IT-IN2P3), \\
Universit{\'e} de Toulouse, CNRS, UPS, F-31062 Toulouse Cedex 9, France.
}

\date{\small \today}

\begin{abstract}

The recent  evidence of a stochastic background of gravitational waves in the nHz band by pulsar-timing array (PTA) experiments has shed new light on the formation and evolution of massive black hole binaries with masses $\sim 10^8$--$10^9 M_\odot$. The PTA data are consistent with a population of such binaries merging efficiently after the coalescence of their galactic hosts, and presenting masses slightly larger than previously expected. This momentous discovery calls for investigating the prospects of detecting the smaller ($\sim 10^5$--$10^7 M_\odot$)  massive black hole binaries targeted by the Laser Interferometer Space Antenna (LISA). By using semi-analytic models for the formation and evolution of massive  black hole binaries calibrated against the PTA results, we find that LISA will observe at least a dozen and up to thousands of black hole binaries during its mission duration. The minimum number of detections rises to $\sim 70$ if one excludes models that only marginally reproduce the quasar luminosity function at $z=6$. 
We also assess LISA's parameter estimation capabilities with state-of-the-art waveforms including higher modes and realistic instrumental response, and find that the masses, sky position, and distance will typically be estimated to within respectively 1\%, 10 square degrees, and 10\% for the detected systems (assuming a 4-year mission).
\end{abstract}


\maketitle

\section{Introduction}

The possible detection of a stochastic background of gravitational waves (GWs)  by the European Pulsar Timing Array (EPTA) \cite{Antoniadis:2023ott}, the Indian Pulsar Timing Array (InPTA) \cite{Tarafdar:2022toa}, the North American Nanohertz Observatory for Gravitational Waves (NANOGrav) \cite{NANOGrav:2023gor}, the Parkes Pulsar Timing Array (PPTA) \cite{Reardon:2023gzh} and the Chinese Pulsar Timing Array (CPTA) \cite{Xu:2023wog}  in June 2023 opened a way for a deeper understanding of GW sources in the nHz frequency band. 
Various exotic explanations have been proposed to account for the origin of the observed stochastic background. These explanations encompass concepts like ultralight dark matter~\cite{EuropeanPulsarTimingArray:2023qbc}, cosmic strings~\cite{EPTA:2023hof}, or cosmological background~\cite{Antoniadis:2023xlr,NANOGrav:2023hvm, Figueroa:2023zhu}. However, the most plausible possibility, by far, is that the pulsar timing array (PTA) signal is produced by an astrophysical population of merging massive black hole (MBH) binaries~\cite{Antoniadis:2023xlr,NANOGrav:2023hfp}.

MBHs, with masses ranging from $\sim 10^5 M_\odot$ (or even lower) to $\sim 10^9 M_\odot$, are ubiquitous in massive galaxies~\cite{Gehren84, Kormendy1995} and in a fraction of low-mass dwarf galaxies~\cite{reines11, reines13, Baldassare2019} in the local universe.  
The evolution of these black holes is intricately connected to that of their galactic hosts, from which they accrete matter, and it is believed that they exert feedback, either through radiation or jets, when they are active galactic nuclei (AGNs) and shine brightly~\cite{Croton2006}. This synergistic evolution is reflected in a wealth of scaling relations~\cite{Kormendy2013, Mcconnell2013, Schramm2013} between the MBH properties and those of their galactic hosts, and is fundamental for galaxy formation models. Indeed, AGN feedback is probably responsible, at least in part, for the occurrence of these scaling relations (see however Refs.~\cite{VolonteriReines16, Shankar2016, Barausse2017, greene19}  for some complications in the interpretation of these relations). Moreover, it is also crucial to reconcile the bottom-up hierarchical formation predicted by the $\Lambda$CDM model for dark matter halos with the ``anti-hierarchical'' evolution (or ``downsizing'') of galaxies~\cite{1996AJ....112..839C,2009MNRAS.397.1776F}, i.e. the fact that massive galaxies mostly contain old stellar populations and show little star formation, while low-mass galaxies are typically dominated by young stellar populations and enjoy more vigorous star formation.

 Despite their importance for galaxy formation, MBHs are still poorly understood, especially at the low end of their mass spectrum and at high redshift, where electromagnetic observations are difficult. GW observatories are crucial in this respect, as gravitational signals decay slowly with redshift (scaling with the inverse of the luminosity distance $D_L$) and interact weakly with matter. In this context, it is clear that the PTA detection of a stochastic GW background is potentially ripe for consequences for our understanding of MBHs. Indeed, as stressed in Ref.~\cite{Antoniadis:2023xlr,NANOGrav:2023hvm,NANOGrav:2023hfp}, the PTA signal is in broad agreement with expectations from our current understanding of the formation and evolution of MBH binaries, provided that the latter merge (and possibly accrete) rather efficiently. However, as the MBHs responsible for the PTA background have masses $\gtrsim 10^8 M_\odot$, different experimental facilities are needed to explore the low end of their mass function, down to $10^5 M_\odot$ or smaller. Among the latter, the Laser Interferometer Space Antenna (LISA)~\cite{LISA2017}, a joint Euro-American space mission, will play a major role.

 LISA will target MBH mergers with masses between $10^4M_\odot$ and
 $10^7M_\odot$, up to very high redshifts ($z\sim 15$ or even larger), and will therefore greatly enhance our understanding of the assembly of MBHs in their galactic hosts~\cite{LISA2017}. Numerous attempts have been made at predicting the number of MBH mergers that will be observed by LISA, based on semi-analytic galaxy formation models and hydrodynamic simulations~\cite{Volonteri2003,Sesana2007b, Sesana2011a, Plowman2011, Rodriguez-Gomez2015, Klein2016, Ricarte2018b, 2019MNRAS.486.4044B, 2019MNRAS.486.2336D,tremmel}, but these estimates are plagued by large uncertainties due to the lack of resolution of the models/simulations and most of all to our incomplete understanding of the ``subgrid'' physics that regulates the evolution of MBHs and galaxies on small scales. The PTA detection, therefore, offers a unique opportunity to test the models used thus far to predict LISA event rates and decrease the error bars of the latter~\cite{Steinle:2023vxs}. This will be the subject of this paper.

 In more detail, the paper is structured as follows: In Section II, we delve into the physics of semi-analytic models for the formation and evolution of galaxies and MBHs. In Section III, we compare the predictions of these models with the data obtained from the PTA observations. Moving on to Section IV, we provide a comprehensive description of GW emission and detection with LISA. Section V is dedicated to presenting our findings concerning LISA's detection rates and parameter estimation. Finally, in Section VI, we summarize our study and draw our conclusions based on the results obtained.

\section{The semi-analytic model}
\label{sec:sam}

We adopt the semi-analytic model of \cite{Barausse2012} for the formation and evolution of galaxies and MBHs.
Additions and improvements to specific aspects of the model were subsequently
introduced in 
\cite{Sesana2014}, \cite{Antonini2015}, \cite{Bonetti2018b}, \cite{2019MNRAS.486.4044B} and \cite{tremmel}. 
Here, we review concisely the model's framework, referring the interested reader to 
the aforementioned works for more details.

\subsection{The dark matter merger trees}

The model is constructed upon a merger tree of dark matter, created using an extended Press-Schechter algorithm~\cite{Press1974,Parkinson2008}, modified to reproduce N-body simulation results~\cite{Parkinson2008}. 
The resolution of the merger tree (i.e. the minimum halo mass below which
dark matter growth via mergers is not followed in detail but collectively modeled as accretion) is redshift dependent. In more detail, it is set to  a fixed fraction of the mass of the most massive halo at the previous (later) redshift step of the tree\footnote{ This fixed fraction
was chosen to be $10^{-2}$ in Ref.~\cite{Barausse2012} and $3\times 10^{-3}$
in the later works using our model, e.g. Refs.~\cite{Sesana2014,Antonini2015,Klein2016,Bonetti2018b,2019MNRAS.486.4044B,tremmel}.}. The redshift step is chosen adaptively
to be sufficiently small
to ensure that multiple halo fragmentation is unlikely~\cite{Volonteri2003}.
A procedure to account for the finite merger tree resolution and extrapolate
MBH merger rates
to infinite resolution was put forward in Fig.~4 of \cite{Klein2016}, and will be adopted below whenever explicitly mentioned.

\subsection{The extrapolation to infinite resolution}
\label{sec:sam:extr}

More precisely, the extrapolation procedure for the (intrinsic) MBH merger rates works as follows. In our simulations for low-mass halos, we notice a linear correlation between
a halo's mass at $z=0$ ($M_0$) and the number of MBH mergers taking place in the past history of that halo. This trend
is easily understood as follows. If one assumes that seed black holes form in halos of mass $M_s$,
the halo at $z=0$ will have formed from $N\approx M_0/M_s$ seed halos. By defining $N\approx 2^n$ and considering a perfect hierarchy
where mergers proceed in subsequent rounds, the total number of mergers by $z=0$ is $\sum_{i=0}^{n-1} 2^i=2^n-1\approx M_0/M_s$, which explains the linear trend
at low masses. At high masses $\gtrsim 10^{13} M_\odot$, this trend is lost, presumably due to lack of resolution. We therefore correct for it by appropriately re-weighing the merger rates
in those halos. 

\subsection{The baryonic structures}

The evolution of baryonic structures  along the merger tree is followed using semi-analytic prescriptions. Among these baryonic structures is a chemically unprocessed intergalactic medium, which accretes onto dark matter halos and undergoes shock heating to the virial temperature in low-redshift, high-mass systems, or which flows into halos along cold filaments on a timescale comparable to the dynamical time in higher-redshift and/or lower-mass systems~\cite{Dekel2006,Cattaneo2006,Dekel2009,correa}.

The intergalactic medium, cooling or streaming along cold filaments into halos, gives rise to a cold gas medium that eventually forms stars (and which is therefore known as  ``interstellar medium''). The model follows the evolution of 
the interstellar medium and the stellar population 
in disks and bulges (i.e. spheroids), accounting for 
the disruption of disks as a result of major galaxy mergers and bar instabilities, as well as for supernova (SN) feedback and the interstellar medium's chemical evolution.

In galactic nuclei, the model forms nuclear star clusters via in-situ star formation and/or  migration of globular clusters~\cite{Antonini_Barausse2015,Antonini2015}. MBHs are also included in the model, forming from seeds at high redshifts and then growing 
by accretion of nuclear gas and mergers with other black holes when galaxies coalesce. 
MBH accretion and nuclear star cluster in-situ formation is assumed to occur from a 
nuclear gas reservoir, whose growth is  modeled as
 linearly correlated with star formation in the bulge~\cite{Granato2004,Barausse2012,Lapi2014,Ricarte2019}. 
The gas in this nuclear reservoir is then assumed to accrete onto the central black hole
 on a viscous timescale evaluated (at the black hole influence radius)~\cite{Sesana2014}, but 
 the accretion rate is capped at  a maximum rate comparable to the Eddington rate (or slightly larger for the light-seed scenario, see below).
  Starting from Ref.~\cite{tremmel}, the model started accounting also for a possible effect of SN feedback
 on the formation of the nuclear gas reservoir. Indeed, the simulations of \cite{Habouzit2017}
show that in SN feedback models where gas cooling is delayed (as a result of shocks), 
accretion is quenched  in low-mass systems. The model of Ref.~\cite{tremmel} therefore
suppresses the reservoir growth in bulges with escape velocities $\lesssim 270$ km/s (i.e. the typical speed of SN winds).
The model also accounts for  AGN feedback from MBHs onto the bulge interstellar  medium and the diffuse chemically pristine gas, both from radio jets~\cite{Barausse2012} and radiation~\cite{tremmel}.

\subsection{The black hole seeds}

A crucial aspect of the model for predicting the event rate for LISA is the initial mass function of the black hole seeds at high redshift.
Several physical models for the latter have been proposed,  see e.g. \cite{Latif:2016qau} for a review. Here, we adopt two
representative scenarios: 
a light-seed (LS) mass function, assuming that
seeds form from the remnants of population III stars  in high-redshift, low-metallicity systems~\citep{Madau2001}; and a
heavy-seed (HS)  mass function, assuming that the seeds arise from the collapse of proto-galactic disks induced by bar instabilities~\citep{Volonteri2008}.
In more detail, in the LS 
scenario we draw 
the  mass of the         
initial population III stars from a log-normal distribution 
centered on $300 M_\odot$ and with a standard deviation of 0.2 dex, but 
with a pair-instability gap between 140 and 260  $M_\odot$~\cite{Heger2002}. We then assume that the black hole seed
has mass $\sim 2/3$ of the mass of the         
initial population III stars, to account for mass losses during the stellar collapse.
Following Ref.~\cite{Volonteri2003}, 
we assume that LSs form only in the 
the most massive halos, corresponding to the 3.5$\sigma$ peaks of the matter density field, at $z\gtrsim 15$. 
Since it is well known that LS models struggle to reproduce the high-redshift luminosity function~\citep{Madau2014}, we allow for mildly super-Eddington  accretion (up to twice the Eddington rate)  in the LS scenario.
In the HS scenario, black hole seeds have masses depending on the properties of the host halo, typically of the order of $\sim 10^5 M_\odot$. 
Several implementations of this scenario have been put forward, with Ref.~\cite{Barausse2012} adopting e.g. the model of Ref.~\cite{2004MNRAS.354..292K}, and later versions of the model (from 2015 onwards) adopting instead the model of  Ref.~\cite{Volonteri2008}. In the latter, in particular, HSs
form from bar instabilities of protogalactic disks at  $z\gtrsim 15$
in halos of masses $\lesssim$ a few times $10^7 M_\odot$. The instability is assumed to occur below a critical Toomre parameter $Q_c$, which is believed to 
be in the range  $2\lesssim Q_c\lesssim 3$. Different choices of $Q_c$ affect the seed occupation fraction at high redshifts, with
lower (larger) $Q_c$ producing fewer (more) seeds. For instance, Ref.~\cite{Bonetti2018b} used $Q_c=2.5$, while
Refs.~\cite{Antonini2015,Klein2016,tremmel} used $Q_c=3$.
It should be noticed that because the HSs of Ref.~\cite{Volonteri2008}
form in relatively small halos ($\lesssim$ a few times $10^7 M_\odot$) at 
high redshift, merger tree branches that fall below the tree's resolution
at low redshifts (and which are therefore not followed to high $z$)
would be artificially devoid of MBH seeds. To alleviate this problem,
in Ref.~\cite{tremmel} our model started following these sub-resolution branches and their dark matter merger history on the fly up to $z\gtrsim 15$, without evolving baryons (for computational efficiency)\footnote{While the correction for sub-resolution branches was already present in Ref.~\cite{tremmel},  unfortunately it was not reported explicitly there.}. If any one of these $z\gtrsim 15$ progenitor halos contains a seed, we place a seed black hole in the sub-resolution branch. Ref.~\cite{tremmel} uses this mitigation strategy for both LSs and HSs, although it is more important for the latter, since our population III seed forms in the largest halos at high $z$. 
We stress that this procedure, like  the other details for the seeding mechanism, has little effect on low $z$ observables, including the PTA stochastic background. This is because, by the time the black holes have
evolved to low redshift or have grown enough to emit in the PTA band, they have lost memory of the initial seeding conditions. However, the seed model, and also our prescription for sub-resolution seeding, has a big impact on LISA merger rates, which are sensitive also to high redshift.
In particular, we will see below that the HS models of Ref.~\cite{tremmel} predict more LISA events than similar earlier models (e.g. those of Ref.~\cite{Klein2016}), and this difference can be ascribed to the updated prescription for sub-resolution seeding.

\subsection{The delays}

Also crucial for the prediction of LISA merger rates is the modeling of the ``delays'' between halo/galaxy mergers
and MBH mergers. Unlike the seeding mechanism at high redshift, these delays also have an important effect on predicting the signal for PTAs~\cite{Bonetti2018b}. 
In more detail, the delays are implemented in our model as follows. When two halos merge according to our dark matter 
tree, we assume that the smaller one survives inside the newly formed system as a satellite subhalo, until it has sunk to the center as a result of dynamical friction~\cite{Boylan-Kolchin2008}. In this phase, which typically lasts a few Gyr, the subhalo, and its baryon content also undergo tidal stripping and evaporation
as a result of the tidal field of the primary, which in turn changes the evolution of the system (decreasing the efficiency of dynamical friction~\cite{Taffoni2003}).

When the subhalo has sunk to the bottom of the host, the baryonic components (i.e. the galaxies) do not merge right away. The satellite galaxy keeps falling towards the center of the host galaxy, again as a result of dynamical friction and tidal stripping/evaporation~\cite{tremmel,Dosopoulou2017,Tremmel2018}.
This phase can last for several Gyr, especially for 
galaxies with unequal stellar masses~\cite{Tremmel2018},
and is crucial to driving the MBHs from $\sim$ kpc to $\sim$ pc separation. However, as a result of tidal stripping and evaporation,
the satellite galaxy may be disrupted and its MBH left as a naked black hole (surrounded at most by a core of stars~\cite{Dosopoulou2017}).
This can lead to a potentially numerous population of  ``stalled'' MBHs   wandering at
separations of hundreds of pc~\cite{Dosopoulou2017,Tremmel2018}.

At separations, $\sim$ pc,  MBHs eventually form
bound binaries, for which dynamical friction becomes inefficient compared to
other processes. These include for instance three-body interactions between the black hole binary and individual stars (``stellar hardening''~\cite{Quinlan1996,Sesana2015}). In more detail, stars with 
low angular momentum (i.e. in the ``loss cone'') 
can remove energy from  the binary via
the slingshot effect. As the process repeats for many stars, the black hole binary shrinks progressively, as stars in the loss cone are
ejected from the nucleus (e.g. as hypervelocity stars~\cite{Sesana2006}). 
This causes hardening eventually to become inefficient unless the 
loss cone gets replenished by stellar diffusion. If this diffusion-induced replenishment
is efficient, e.g. as a result of triaxiality of the galaxy potential (resulting for instance from a recent galaxy
merger)~\cite{Yu2002,Khan2011,Vasiliev2014,Vasiliev2014a,Vasiliev2015} or galaxy rotation~\cite{Holley-Bockelmann2015}, stellar hardening can 
drive the binary separation down to $\sim 10^{-2}$--$10^{-3}$ pc 
on a timescale of few Gyr. Once those small separations have been reached, GW emission alone leads the binary to merger 
 in less than a Hubble time.

Other processes may also help the binary reach sub-pc separations (see e.g. Ref~\cite{Colpi2014} for a review). For instance, in gas-rich galactic nuclei, the MBH binary may shrink on 
timescales of $\sim 10^7$--$10^8$ yr thanks to interactions with the nuclear
gas (planetary-like  migration). Furthermore, even in gas-poor environments
 with inefficient loss-cone replenishment (and thus inefficient stellar hardening), triple MBH systems will eventually form as a result of later galaxy mergers.
 In such triple systems, Kozai-Lidov oscillations~\cite{Kozai1962,Lidov1962}
 and/or chaotic
three-body interactions~\cite{Bonetti2016,Bonetti2018a} can trigger the merger of at least two of the black holes in a sizeable fraction of systems~\cite{Bonetti2018a,Bonetti2018b,2019MNRAS.486.4044B,tremmel}. Interestingly, these triplet-induced mergers are expected to present
 a significant 
eccentricity ($\gtrsim 0.99$ when they enter the LISA band, and $\sim 0.1$ at the merger)~\cite{2019MNRAS.486.4044B}.

\section{Comparison with PTA data}
\label{sec:PTA}

Preliminary implications of the PTA detections for the formation and evolution of MBHs have been presented in Refs.~\cite{Antoniadis:2023xlr,NANOGrav:2023hvm,NANOGrav:2023hfp}. Here, we follow and extend Sec. 3.3.1 of Ref.~\cite{Antoniadis:2023xlr} and perform a comparison between the EPTA measurement of the stochastic background and the predictions of our semi-analytic model.
Like in Ref.~\cite{Antoniadis:2023xlr}, we consider in particular
 the semi-analytic model of
\cite{Barausse2012} in its original version (B12) and in its subsequent updates, which were introduced and used in Refs.~\cite{Antonini2015,Klein2016} (K+16), Refs.~\cite{Bonetti2018b,2019MNRAS.486.4044B} (B+18) and \cite{tremmel} (B+20).

The specific models 
that we consider (which include as a subset those considered in 
Ref.~\cite{Antoniadis:2023xlr}) can be roughly divided in three classes: \textit{(i)}
 ``LS-nod (B12)'', ``HS-nod (B12)'', ``Q3-nod (K+16)'',
``LS-nod-noSN (B+20)'', ``HS-nod-noSN (B+20)'', ``LS-nod-SN (B+20)'', ``HS-nod-SN (B+20)''
and ``HS-nod-SN-high-accr (B+20)'' assume no delays
between galaxy and MBH mergers (although they account
for the delays between  halo and galaxy mergers, c.f. Sec.~\ref{sec:sam})\footnote{Models ``LS-nod-noSN (B+20)'', ``HS-nod-noSN (B+20)'', ``LS-nod-SN (B+20)'', HS-nod-SN (B+20)''
and ``HS-nod-SN-high-accr (B+20)'' 
are produced with the semi-analytic model of B+20, but with no delays between 
galaxy and MBH mergers (except for the dynamical friction timescale -- including tidal effects -- between dark matter halos). These models were 
 {\it not} presented in B+20, but only in Ref.~\cite{Antoniadis:2023xlr}.}; \textit{(ii)}  ``popIII-d (K+16)'', ``Q3-d (K+16)'', ``LS-d (B+18)'', ``HS-d (B+18)'', ``LS-noSN-sd (B+20)'', ``LS-SN-sd (B+20)'', ``HS-noSN-sd (B+20)'' and ``HS-SN-sd (B+20)''
account not only for the delays between halo and galaxy mergers but also for stellar hardening, MBH triplets and planetary-like migration;  \textit{(iii)}  ``LS-noSN-d (B+20)'', ``LS-SN-d (B+20)'', ``HS-noSN-d (B+20)'', ``HS-SN-d (B+20)'', ``LS-inf-d (B+18)'' and ``HS-inf-d (B+18)''
account for the full delays, i.e. they also include the evolution of black hole pairs at
 separations of hundreds of pc (cf. Sec.~\ref{sec:sam}). 
 The labels  ``SN'' and ``noSN'' respectively indicate models that do and do not account for the effect of SN feedback on the growth of the nuclear gas reservoir, while
``LS''/``popIII'' and ``HS''/``Q3'' denote respectively 
LS and HS scenarios for the high redshift initial mass function.

\begin{figure}
    \centering
    \includegraphics[width=0.5\textwidth]{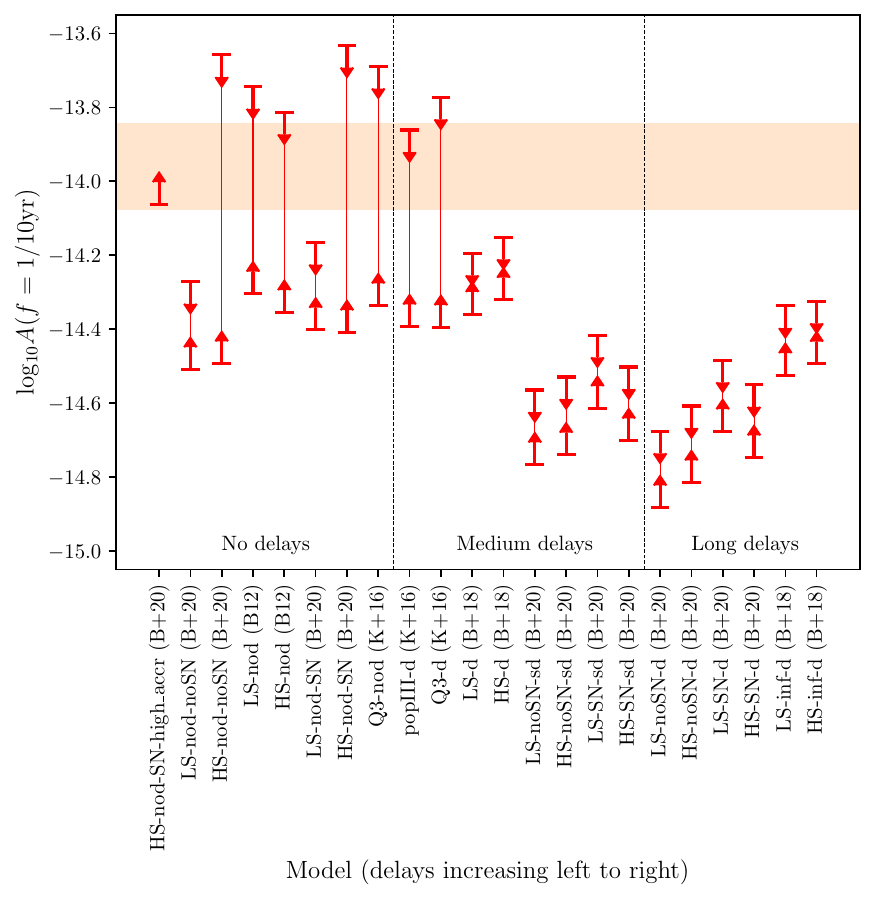}
    \caption{\footnotesize{Characteristic strain amplitude at $f=1/10$yr predicted by various published semi-analytic galaxy formation models, assuming quasi-circular inspiral, no interactions with matter, and without accounting for cosmic variance (i.e. for the scatter among different realizations of the black hole population).
    The models feature
    different physical assumptions for the delays 
    between the galaxy  and black hole mergers    
    (increasing from left to right). 
    The ``no delays'', ``medium delays'' and ``long delays'' models correspond  to the model classes \textit{(i), (ii)} and \textit{(iii)} defined in the text, respectively.   
    For each model,
    the lower bound is the result at finite resolution, while the upper one is the extrapolation to infinite resolution (see text for details).
The shaded area is the 
    95\% confidence region from the EPTA measurement assuming $\gamma=13/3$~\cite{Antoniadis:2023xlr}. This figure is adapted from Fig. 8 of Ref.~\cite{Antoniadis:2023xlr}.}}
    \label{fig:sam}
\end{figure}

Each model's predictions are compared to the EPTA measurement in Fig.~\ref{fig:sam}, which is adapted from Fig. 8 of Ref.~\cite{Antoniadis:2023xlr} and which we report here for completeness. The model predictions assume quasi-circular orbits for the MBH binaries, as a result of which the spectrum has a slope  $\gamma=13/3$,
and are obtained by summing the
 GW energy spectra of the whole theoretical
binary population. For this reason, the predictions do not have any cosmic variance, i.e. at this stage we do not consider the scatter  among different realizations of the binaries in our past light cone. The shaded area is the EPTA measurement of the amplitude at $f=1/10\,$yr assuming  $\gamma=13/3$.

Note that each model's prediction [except model HS-nod-SN-high-accr (B+20)]
is shown as a range. The lower end represents the prediction of finite merger tree resolution, while the upper end is the extrapolation to infinite resolution. This extrapolation is performed following Fig. 4 of  \cite{Klein2016}  (i.e., as described in Sec.~\ref{sec:sam:extr}), but because of the uncertainties involved, it should be considered as an upper limit.
For  model HS-nod-SN-high-accr (B+20) we report only the (more robust)  finite resolution prediction, which
agrees already with the measurement. This agreement is the result of  stronger black hole accretion, which is obtained in this model by boosting the influx of gas into the nuclear reservoir during star formation events in the spheroid by a factor
 $\sim 4$.

\begin{figure}
    \centering
    \includegraphics[width=0.5\textwidth]{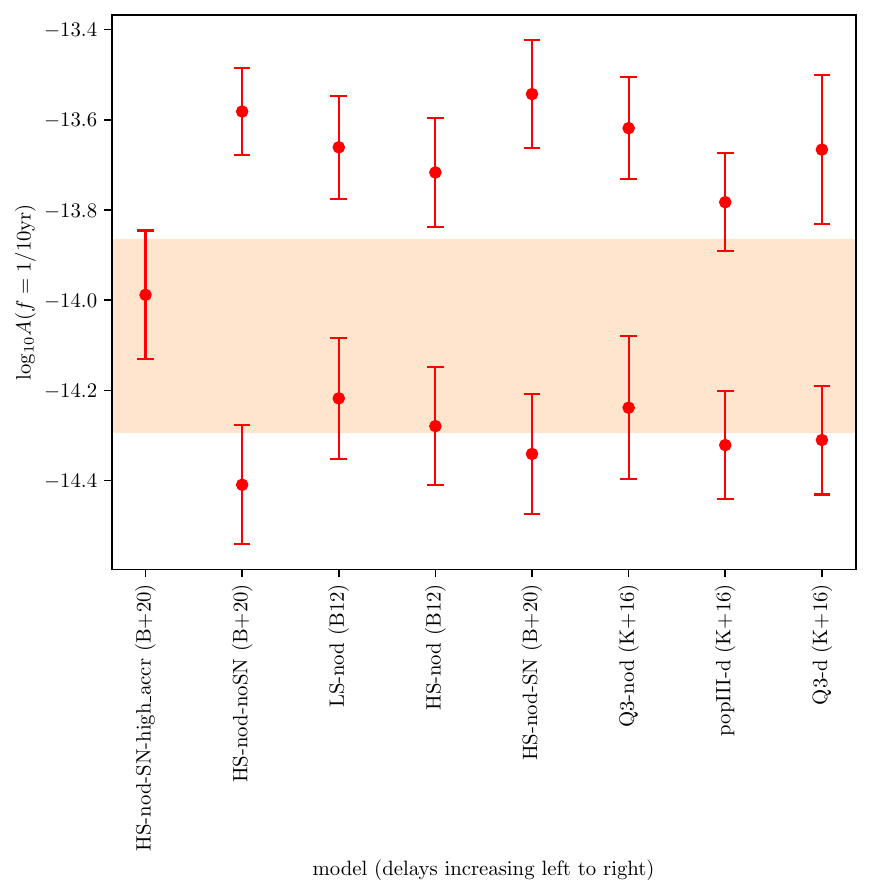}
    \caption{\footnotesize{
    Predicted characteristic strain amplitude at $f=1/10{\rm yr}$ from various semi-analytic galaxy formation models. The amplitude is obtained by producing multiple realizations 
    of the population of MBH binaries, and then by fitting the resulting spectrum in the first nine frequency bins (assuming an observation time of 10.3 yr) with a power law. The error bars account for the scatter among the different realizations (cosmic variance) and represent  95\% confidence intervals. For each model, 
    we present both the result at finite resolution (lower) and the extrapolation to infinite resolution (higher), except for model
  HS-nod-SN-high-accr (B+20) (finite resolution only).
      The shaded area is the 95\% confidence region from the EPTA measurement (marginalizing over the spectrum's slope)~\cite{Antoniadis:2023xlr}.}
    \label{fig:sam3}}
\end{figure}

In Fig.~\ref{fig:sam3}, we consider only the models in better agreement with the data.\footnote{Given the uncertainties in the PTA measurements and the fact that we assume quasi-circular binaries in our predictions, we consider all models in Fig.~\ref{fig:sam3} to be in broad agreement with the PTA results.} Here, we do not sum the spectra of the whole theoretical population, but we consider multiple realizations of the binaries in our past light cone, following Ref.~\cite{Sesana_Vecchio2008}. We then fit each realization's predicted spectrum with a power law in the first 9 frequency bins and report the average signal amplitude 
 $A(f=1/10{\rm yr})$
 and its 95\% confidence region. Note that this region (i.e. the error bars) should  therefore be interpreted as representing cosmic variance.
These predictions are compared to the corresponding range of the measurement, i.e. the 95\% confidence region of the amplitude $f=1/10\,$yr, but this time we marginalize over the slope $\gamma$ (cf. Fig. 1 of Ref.~\cite{Antoniadis:2023xlr}), since the model predictions do not have $\gamma=13/3$ as a result of the finite number of realizations~\cite{Sesana_Vecchio2008}.

As stressed in Ref.~\cite{Antoniadis:2023xlr}, this comparison allows one to draw the following qualitative conclusions:
{\it (i)} large delays at separations of hundreds of pc are disfavored and  MBHs merge efficiently after galaxy mergers; 
{\it (ii)} accretion onto MBHs may be more efficient
than previously thought, resulting potentially in a larger
local MBH mass function  at the high-mass end.

\begin{figure*}
    \centering
    \includegraphics[width=\textwidth ]{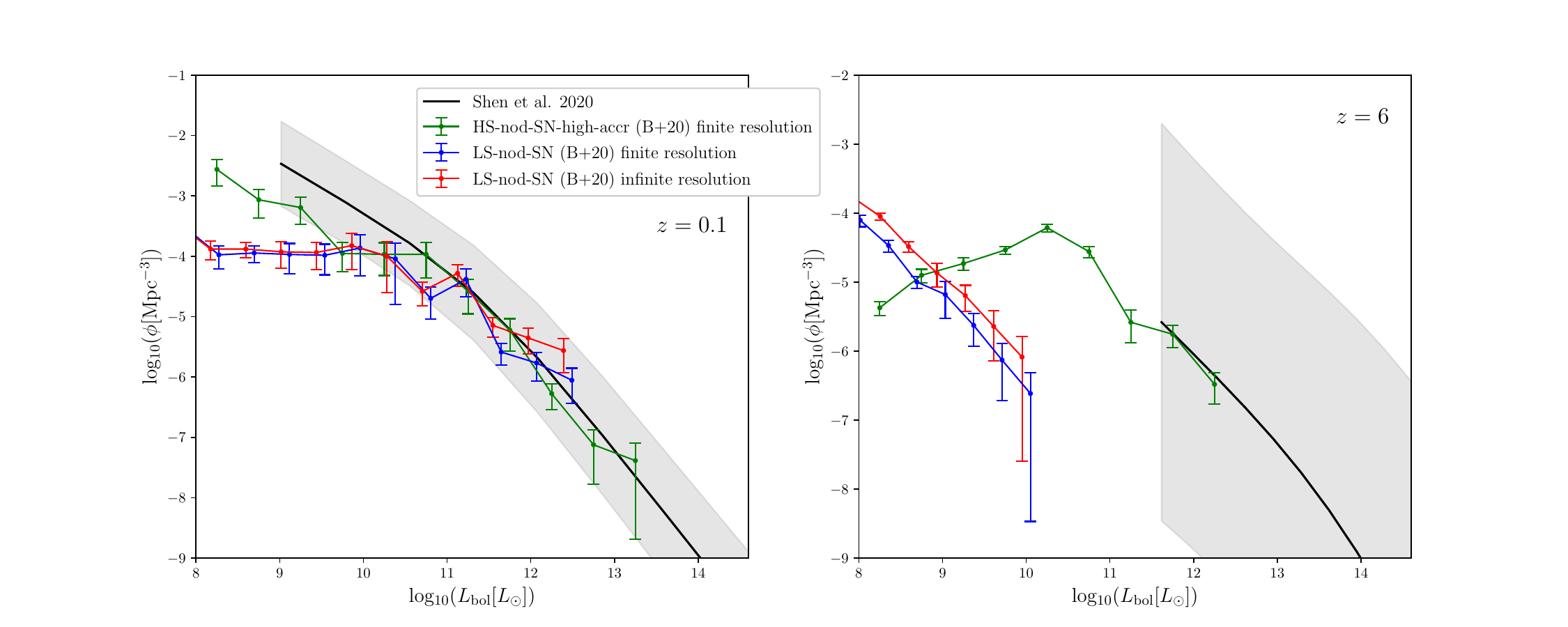}
    \includegraphics[width=\textwidth]{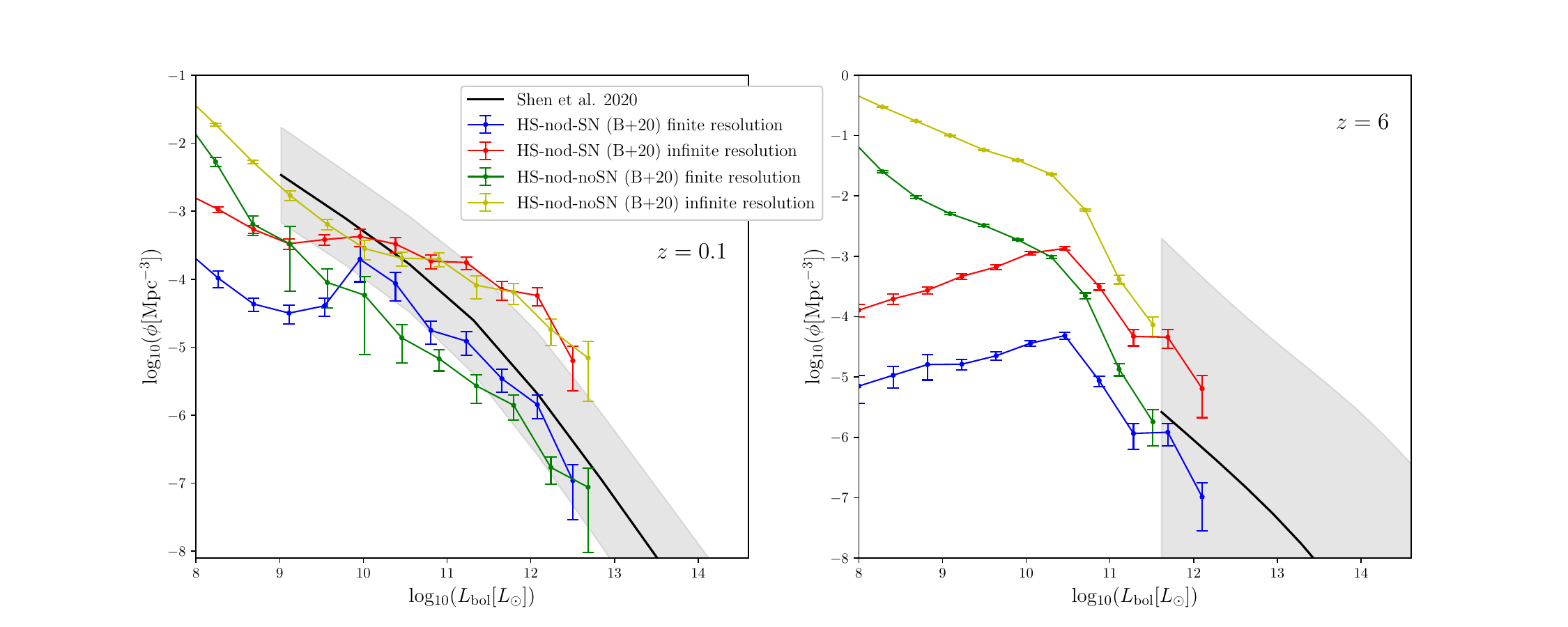}
    \caption{\footnotesize{Bolometric quasar luminosity function at $z=0.1$ (left panels) and $z=6$ (right panels) for different models described in the text. The black line with shaded $95\%$ confidence interval is the observed bolometric luminosity function  from Ref.~\cite{2020arXiv200102696S}.}}
    \label{fig:LF}
\end{figure*}

To check that the boost in the reservoir growth rate in model HS-nod-SN-high-accr (B+20) did not result in an overprediction of the local bolometric quasar luminosity function, in the upper panels of Fig.~\ref{fig:LF} we compare 
to data reported in Ref.~\cite{2020arXiv200102696S}, which show good agreement.\footnote{Note that  it is not obvious that our boosted accretion model will agree with the local luminosity function. In our case, it does agree because of our
treatment of the radiative efficiency at high accretion rates (cf. Eq. 40 of B12, which approximately represents the transition from thin to slim disk accretion as the mass accretion rate increases).} 
Similarly, because it is expected~\cite{Madau2014,tremmel} that LS models can only reproduce the luminosity function of quasars at $z=6$ in the presence of a sustained phase of super-Eddington accretion, in the same panels we also show results from LS-nod-SN (B+20). As can be seen,  model LS-nod-SN (B+20) struggles to reproduce the observed luminosity function as a result of SN feedback, as also found in Ref.~\cite{tremmel}. We will include model LS-nod-SN (B+20) when making predictions for LISA, but one should keep in mind that it is marginally disfavored by the $z=6$ quasar luminosity function.
In the lower panels of Fig.~\ref{fig:LF}, we show results from models HS-nod-noSN (B+20) and HS-nod-SN (B+20) for comparison.

\section{GW emission, detection and parameter estimation}
\label{sec:GW}

We produce synthetic catalogs of MBH binaries to characterize their emitted GW signals. Our main goal is to assess their detection rate and the parameter estimation prospects for LISA. 
We describe the GW signal from an MBH binary  by a set of 11 parameters - the masses $M_1$ and $M_2$, the (aligned) spins $\chi_1$ and $\chi_2$, the time of coalescence $t_c$, the luminosity distance $D_L$, the inclination of the angular momentum of the source relative to the line of sight $\iota$, the sky longitude and latitude $\lambda$ and $\beta$, the polarization angle $\psi$ and the phase at coalescence $\phi$. If one were to include precession in the analysis, there would be extra parameters describing the orientation of the spins. In this study, we employ IMRPhenomHM \cite{London:2017bcn}, which is a non-precessing waveform model that describes binaries with aligned spins\footnote{{ While a recent work~\cite{Pratten:2023krc} included both precession and higher modes in the analysis of MBH binaries observable by LISA, we focus here on the non-precessing case for simplicity. It should be noted, however, that precession can play a crucial role in binary evolution. This will be addressed in a future work.}}, but which  includes contributions from higher order modes. The inclusion of higher harmonics helps measure the source parameters, and particularly  the luminosity distance. 

The signal for LISA has two main components - the GW waveform from the source and the LISA response function. In the frequency domain, the waveform takes the form $h_{\ell m} = A_{\ell m}(f)e^{-i \Psi_{\ell m}(f)}$. The single-arm interferometric measurement, which describes the shift in frequency of the laser linkage $l$ between the spacecraft's $s$ and $r$, is 

\begin{equation}
    y_{slr} = \sum_{\ell ,m}\mathcal{T}_{slr}^{\ell m}(f)h_{\ell m} \,,
\end{equation}
where $\mathcal{T}_{slr}^{\ell m}$ is the transfer function for the $\ell m$ mode. A computation of $\mathcal{T}_{slr}$ without any approximations, although possible, is unfeasible for parameter estimation due to its high computational cost. A fast method of calculating it using a perturbative approach was extended to merger-ringdown in Ref.~\cite{Marsat:2018oam}, making the computation suitable for use in parameter estimation. 

Laser noise dominates the single link observable $y_{slr}$. However, Time Delay Interferometry (TDI) \cite{Estabrook:2000ef,Armstrong_1999,Dhurandhar:2001tct,Tinto:2004wu}, which relies on three independent time-delayed linear combinations (referred to as A, E, and T channels) of 
the single link outputs,  removes the laser noise. The T channel has very low signal content and therefore is not included in our analysis, i.e. we will only work with the A and E channels.

We simulate  gravitational waveforms for each binary in our synthetic catalogs, neglecting the fact that the true LISA data will contain overlapping signals (i.e. we consider each source ``in isolation''). We assume   noiseless simulated data when computing signal-to-noise ratios (SNRs) and parameter estimation. We also neglect possible data gaps and glitches \cite{Dey:2021dem, Baghi:2019eqo, Spadaro:2023muy}, but include the LISA response function. The inclusion of a Gaussian noise realization in the simulated data is expected to shift the posterior distribution up to the statistical error but is unlikely to change the variance of the posterior distribution \cite{Rodriguez:2013oaa, Sampson:2013lpa, Nissanke:2009kt}.

To assess the detectability of a GW source with LISA, we compute its SNR $\rho$, defined as 
\begin{equation}
    \rho^2 = (h|h) \,,
\end{equation}
where $h$ is the GW signal, and the inner product $\left(h|h\right)$ is defined as
\begin{equation}
    (a|b) = 4 \text{Re} \int_0^\infty \frac{\Tilde{a}^*(f)\Tilde{b}(f)}{S_n(f)}\text{d}f \,,
\end{equation}
where $S_n(f)$ is the noise power spectral density.
For the latter, we use the \texttt{SciRDv1} noise model \cite{LISA_SciRDv1} with the addition of an unresolved white dwarf background from Galactic binaries \cite{Babak2017, Cornish2018}. 
As we have higher modes in the waveform, the calculation of $\rho^2$ results in  non-negligible cross terms between different modes (Eq. (26) of Ref. \cite{Pitte:2023ltw}). Moreover, the total  SNR is the sum (in quadrature) of the contributions of two independent TDI observables A and E. 

Each detection (with $\rho\geq8$)  is followed by a full parameter estimation. This provides information on the number of sources for which we can estimate the posterior distribution with errors lower than a specified threshold. 
According to  Bayes' theorem, the posterior distribution of the parameters $\theta$ given the data $d$ is provided by
\begin{equation}
    p\left(\theta|d\right) = \frac{\mathcal{L}\left(d|\theta\right)p\left(\theta\right)}{p\left(d\right)} \,,
\end{equation}
where $\mathcal{L}(d|\theta)$ is the likelihood, $p(\theta)$ is the prior and $p(d)$ is the evidence. The evidence acts as a normalization factor and is useful when comparing different models. While sampling we choose the following parameters - total mass $M$, mass ratio $q$, $\chi_1$, $\chi_2$, $t_c$, $D_L$, $\iota$, $\lambda$, $\beta$, $\psi$ and $\phi$. We choose uniform priors on $M$, $q$, $\chi_1$, $\chi_2$, $t_c$, $D_L$, $\cos\iota$, $\psi$ and $\phi$. For the sky position, we choose uniform priors on $\lambda$ and $\sin\beta$.

The likelihood is computed using the \texttt{LISAbeta} code described in \cite{Marsat:2020rtl}. To sample the posterior distribution, we use \texttt{ptemcee} \cite{2016MNRAS.455.1919V} which implements a parallel-tempered ensemble sampler. Parallel tempering \cite{PhysRevLett.57.2607,B509983H} is a variation of the commonly used Markov Chain Monte Carlo (MCMC) based techniques, and it is efficient at sampling complex posterior surfaces with multiple sharp peaks. As the source parameters of the injected data are known, and our aim is to estimate the parameter errors from the posterior distribution, we can further help the sampler by initializing it at the injection's ground true, with a preliminary covariance of the parameters from the Fisher matrix. The Fisher matrix elements are calculated as
\begin{equation}
    F_{ij} (\theta_{\text{inj}}) = \left(\partial_i h | \partial_j h\right)\big|_{\theta_{\text{inj}}} \,,
\end{equation}
where $\partial_i$ is the derivative with respect to the $i^{\text{th}}$ element in the parameter set $\theta$, and $\theta_{\text{inj}}$ is the injection's ground truth. Initializing the sampler around the true value of the parameters, and using the Fisher matrix in the preliminary stage, we can focus directly on the regions with high posterior probability, without having to sample the full parameter space. The chains, therefore, converge faster, leading to lower computational costs.

This approach, unfortunately, will ignore secondary peaks in the posterior distribution, if proposal distributions are not tuned accordingly.  MBH signals are known to present degeneracies among  sky position, inclination, and polarization \cite{Marsat:2020rtl}. We, therefore, choose the proposal distributions  to allow the sampler to jump between these degenerate modes. 



\section{Results}
\label{sec:results}

\subsection{LISA detection rates}

For detectability, we consider an SNR threshold of 8. Predictions for the number of LISA detections for a 4-year mission are listed in Table~\ref{tab:lisarate}, for the different models. As can be seen, LS models predict fewer detectable sources than HS ones, 
because lower BH masses lead to lower SNRs.
In more detail, lower-mass sources  merge at higher frequencies. Although LISA is nominally sensitive in the band $\sim$0.01 mHz to $\sim$0.1 Hz, at frequencies close to these boundary values the noise dominates, as can be seen from the noise power spectral density {\bf~\cite{LISA_SciRDv1}}. As the merger is the loudest part of an MBH binary signal, missing it results in lower SNR. In fact, for very low masses $\sim 10^3 M_\odot$, the merger may even be completely outside the LISA sensitivity band. This effect is exacerbated by the effect of SN feedback, which may hamper accretion in low-mass/high-redshift galaxies (where SN winds can exceed the typical escape velocities)~\cite{Habouzit2017,tremmel}. While this lack of accretion growth does not prevent binaries from radiating in the LISA band in HS models (where the seed masses already correspond to frequencies LISA is sensitive at), in LS models it leads to very low binary masses at high $z$. As a result, the LS-nod-SN (B+20) model is  the one with the lowest number of events. However, as mentioned above, this model is in marginal tension with the observed quasar luminosity function at $z=6$.
Also noticeable, in Table~\ref{tab:lisarate}, is the difference in the number of detections between the models of B12 and K+16\footnote{ We stress that the detections rates computed for the models K+16 are consistent with those reported in Ref.~\cite{Klein2016} for the N2A2M5L6 configuration, once
one accounts for \textit{(i)} the different mission duration (5 yr in  Ref.~\cite{Klein2016} vs 4 yr here); \textit{(ii)} the slightly different detector configuration (2 Gm arms
in  Ref.~\cite{Klein2016} vs 2.5 Gm here); and \textit{(iii)} the inclusion of the full LISA response and higher modes (both absent in  Ref.~\cite{Klein2016}), which are particularly important for LS models.} and
the later models of B+20. This is due to the updated procedure to resolve early mergers in sub-resolution branches of the merger tree, which we
discussed in Sec.~\ref{sec:sam} and which was introduced in the models of B+20.
An exception is model HS-nod (B12), which predicts a large number of detections
because it adopts a different seeding scenario, namely that of Ref.~\cite{2004MNRAS.354..292K}.
The results of B12 and K+16 can be therefore    
viewed as conservative.

Note also that the impact of the extrapolation to infinite resolution depends on the seeding model and the distribution of the MBH mergers. For instance, our HS  models
are more affected by finite resolution effects, because the seeds can form in relatively small halos~\cite{Volonteri2008}, while LSs are assumed to form in the largest halos at high redshift~\cite{Barausse2012,Madau2001}. Similarly, no-delay models tend to predict more mergers at high $z$, where finite resolution may be an issue. These subtle effects are reflected in the different extrapolated rates in Table~\ref{tab:lisarate}.\footnote{On top of this, the old results of Ref.~\cite{Barausse2012} were produced with slightly lower resolution than the more recent results of 
Refs.~\cite{Sesana2014,Antonini2015,Bonetti2018b,2019MNRAS.486.4044B,tremmel}, as mentioned above.
}

\renewcommand{\arraystretch}{1.5}
\begin{table}
	\begin{tabular}{l|c|c}
		\hline \hline
		\multirow{ 2}{*}{Model} & \quad $N_{\rm det} (4 \ \rm yr)$ \quad& \quad $N_{\rm det} (4 \ \rm yr)$ \quad \\
        & \quad {\it finite res.} \quad & \quad {\it inf. res.} \quad \\
		\hline
		HS-nod-SN-high-accr (B+20) & $8901$ & - \\
		LS-nod-noSN (B+20) & $203 $ & $250 $ \\
		HS-nod-noSN (B+20) &  $15821 $ & $38712 $ \\
		LS-nod (B12) & $432$ & $570$ \\
        HS-nod (B12) & $6154 $ & $7184$ \\
        LS-nod-SN (B+20) & $11$ & $12$ \\
        HS-nod-SN (B+20) & $16133$ & $36090$ \\
        Q3-nod (K+16) & $468$ & $656$ \\
        popIII-d (K+16) & $183$ & $339$ \\
        Q3-d (K+16) & $33$ & $74$ \\
		\hline \hline
	\end{tabular}
    \caption{Number of detections ($N_{\rm det}$), for the different models described in the text, in a 4-year LISA mission.}
	\label{tab:lisarate}
\end{table}

\begin{figure*}
    \centering
    \includegraphics[width=0.49\textwidth]{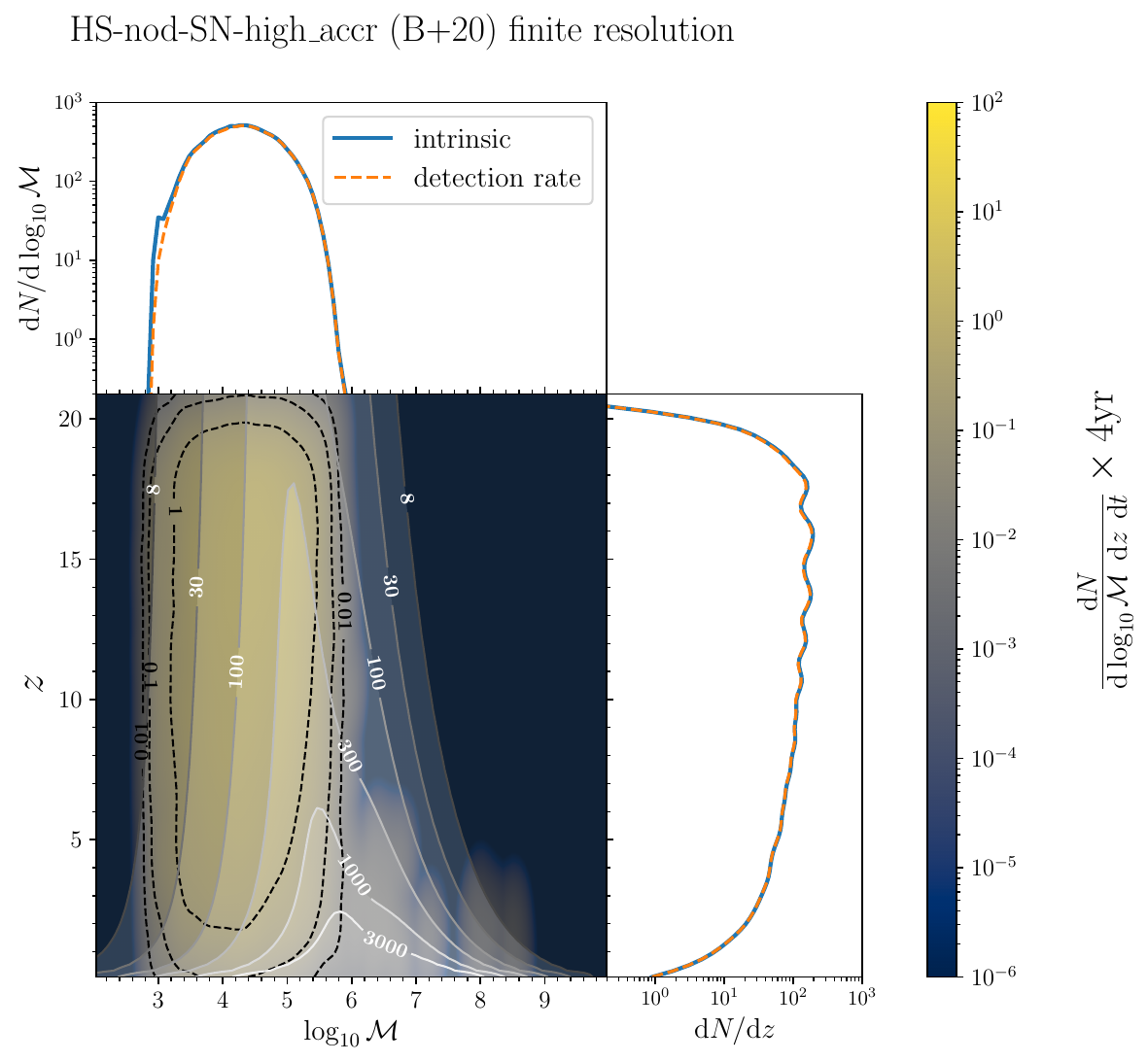}
    \includegraphics[width=0.49\textwidth]{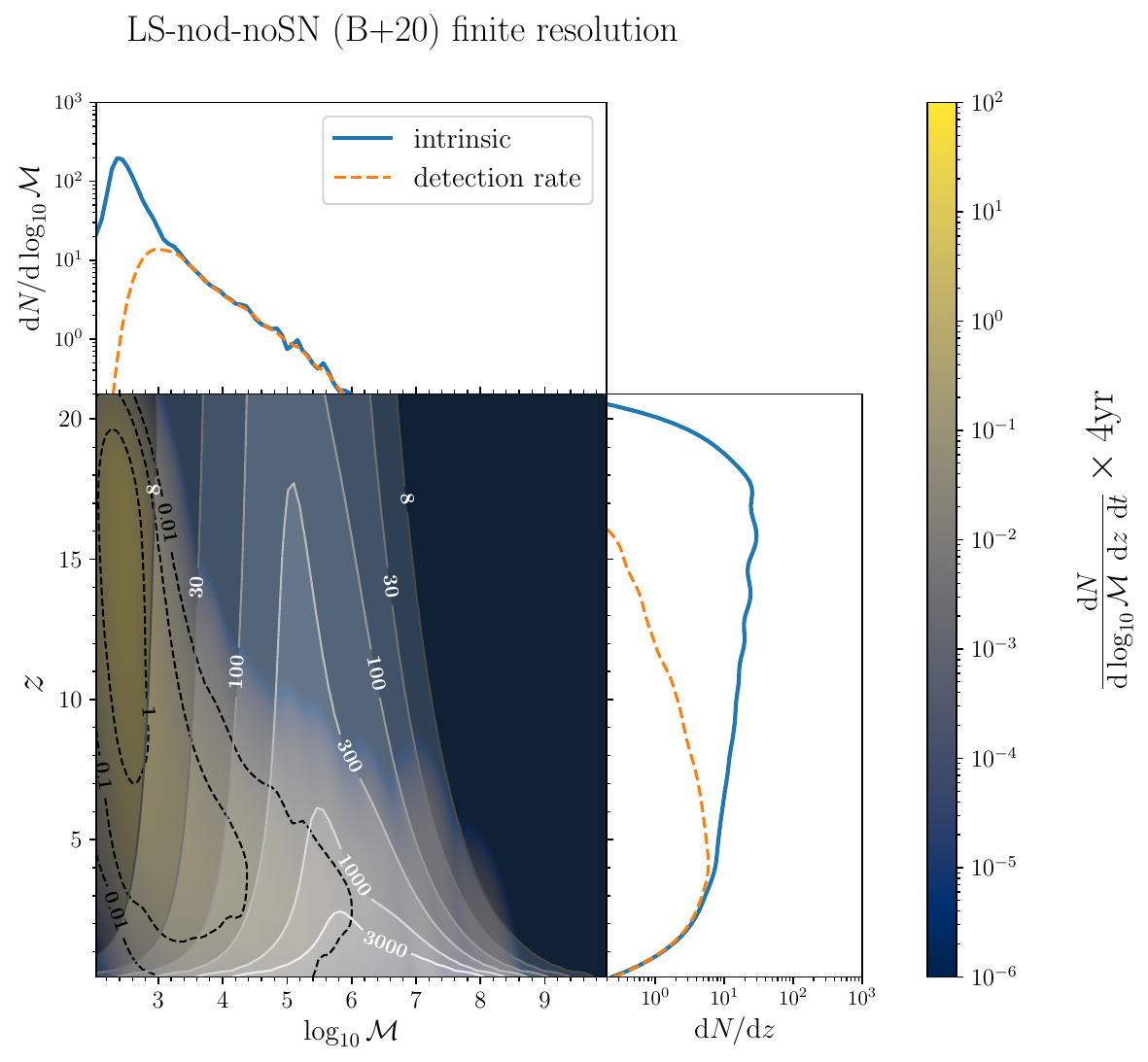}
    \caption{\footnotesize{Number of mergers (color gradient and black dashed lines) in 4 years for models ``HS-nod-SN-high-accr (B+20)'' (left panel) and ``LS-nod-noSN (B+20)'' (right panel), as a function of source-frame chirp mass and redshift. Superimposed (labeled by white numbers), we show the LISA sensitivity in terms of contours of constant SNR (assuming $q=1$). In the upper and right panels, we show the distribution of the total (intrinsic) number of mergers (blue line) and that of the detected number (orange dashed line), marginalizing respectively over redshift and chirp mass.}}
    \label{fig:rate1}
\end{figure*}

To further visualize  the models' physical content and understand their implications for LISA, 
in Fig.~\ref{fig:rate1} we show the  
number of mergers (with no threshold in SNR) in 4 years for models HS-nod-SN-high-accr (B+20) (left panel) and LS-nod-noSN (B+20) (right panel),
as a function of redshift and source-frame chirp mass. Also shown are SNR contours (computed assuming $q=1$), highlighting the parameter space region in which LISA is sensitive. The side panels report marginalized distributions over chirp mass and redshift, for all systems (intrinsic) and the detected ones.
Similar figures are shown in Appendix~\ref{allmergedet} for the other models (at finite resolution; figures at infinite resolution only show
a slightly higher normalization but are otherwise very similar). Again, one can notice that the LS models (and especially those with SN feedback)
lead to very few systems in the region where LISA is sensitive, i.e. only a fraction of the whole population is detectable. In the HS models, instead,
the detection fraction is close to 100\%. Moreover, in the HS models, the detected event have typically larger SNR (up to thousands) than in LS models, again as a result of the higher masses.
These larger SNRs will result in correspondingly low errors on the parameters in the next section, where we will explore the parameter estimation capabilities of LISA.

\subsection{Parameter Estimation}

To assess the LISA parameter estimation capabilities, which are of course crucial to extract science from the data,
we consider the detected sources (with $\rho>8$)
and apply to them our Bayesian pipeline described in Sec.~\ref{sec:GW}.
We run the sampler with 64 walkers and 10 temperatures to produce 8000 samples per detected source. For the calculation of the likelihood integral, we set $f_{\text{min}}=10^{-5}$ Hz and $f_{\text{max}}=0.5$ Hz, which corresponds to the region of the LISA sensitivity curve most favorable to MBH binary detections\footnote{ It is worth noting that these frequency limits indicate the \textit{mission goals}, while the \textit{mission requirement} for the lower bound of the LISA band is $f_{\text{min}}=10^{-4}$ Hz~\cite{LISA_SciRDv1}. This difference in the lower bounds is unlikely to have a significant impact on SNR calculations, as the sensitivity is quite low in the band $10^{-5}-10^{-4}$ Hz. As for parameter estimation, inclusion of higher modes (as we do here) can mitigate potential degradation in the estimated posterior distributions for short-lived signals~\cite{Pratten:2022kug}. Moreover, for the vast majority of signals, which are in the LISA band for longer periods (months or years), the exact choice of the lower frequency bound is unlikely to have an appreciable impact.}. As can be observed in Table \ref{tab:lisarate}, the number of detections is very large for models HS-nod-SN-high-accr (B+20), HS-nod-noSN (B+20), HS-nod (B12) and HS-nod-Sn (B+20); therefore, for these models, we run the parameter estimation pipeline on a subset of sources and extrapolate to the whole detected population.
Model LS-nod-SN (B+20) has the opposite problem, i.e. because it has  a very low detection rate,  to obtain statistically relevant conclusions
we draw our Bayesian inference for
more sources than those that would be detected in
4 years. More precisely, we consider 100 years of data from the synthetic catalogs and rescale the rate   to the LISA mission duration.
\begin{figure}
    \centering
    \includegraphics[width=0.4\textwidth]{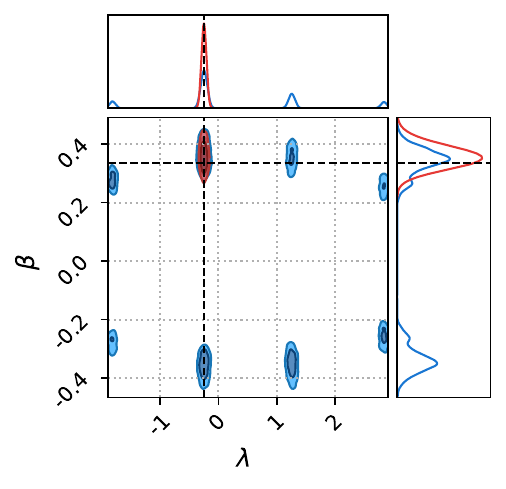}
    \caption{\footnotesize{Multi-modality in the sky position measurement. The ``raw'' posterior distribution without any post-processing is shown in blue. The primary maximum, to which we restrict for error computation, is shown in red. The parameters of this source are - ($M_1 = 2.3\times10^6 M_\odot$, $M_2 = 1.8\times10^5 M_\odot$, $\chi_1=0.03$, $\chi_2=-0.02$, $t_c=2522880$s, $D_L=152$ Gpc, $\iota=2.4$, $\lambda=-0.25$, $\beta=0.33$, $\psi=0.92$, $\phi=-0.64$).}}
    \label{fig:multi-modal}
\end{figure}

\begin{figure*}
    \centering
    \includegraphics[width=\textwidth]{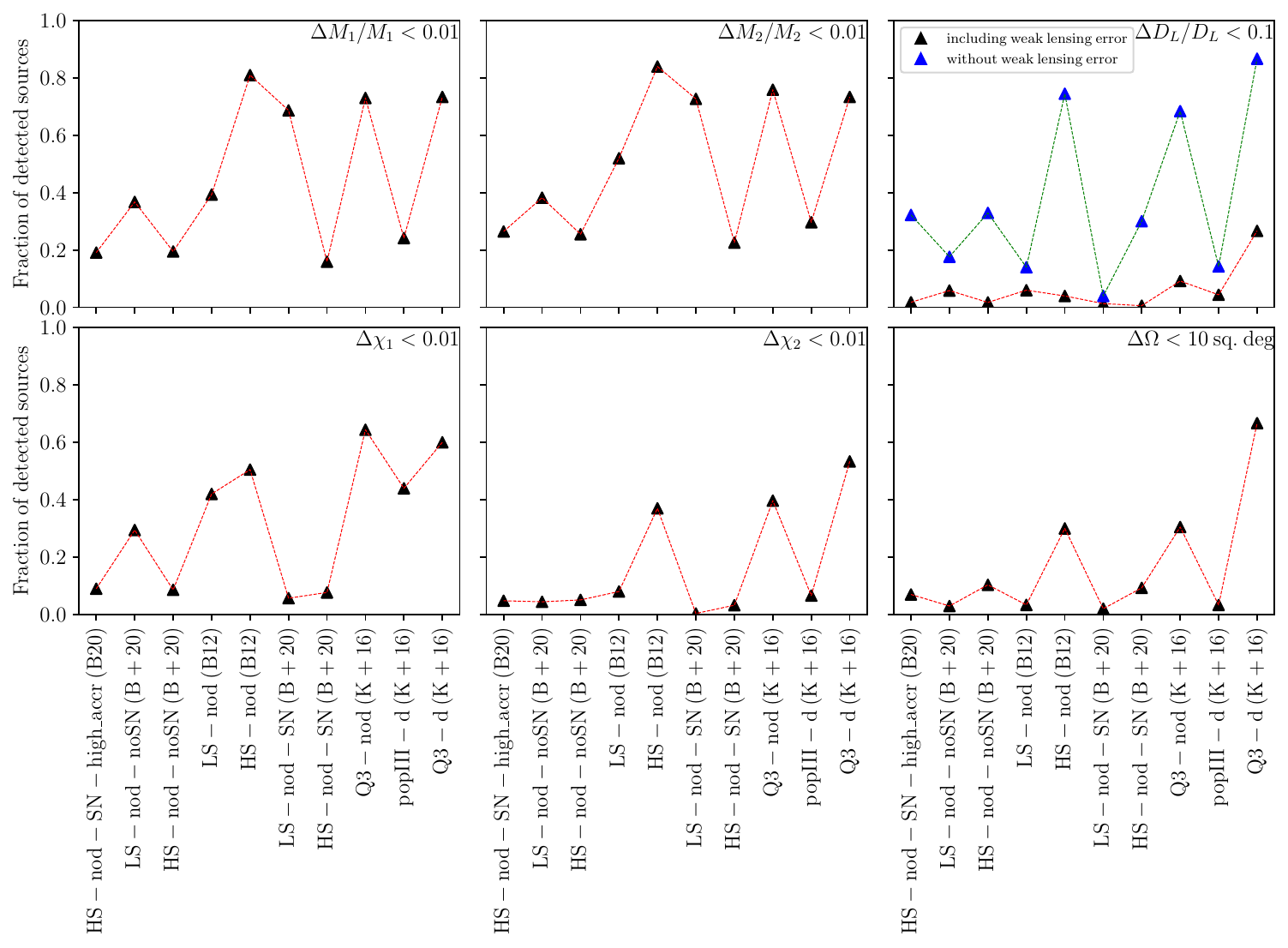}
    \caption{\footnotesize{Fraction of detected sources satisfying 
    given      thresholds on the parameter errors. In more detail, the top panels, from left to right, show sources with $\Delta M_1/M_1< 0.01$, $\Delta M_2/M_2<0.01$, and $\Delta D_L/D_L<0.1$; 
    the bottom panels show sources with $\Delta\chi_1<0.01$, $\Delta\chi_2<0.01$ and
    $\Delta\Omega < 10\:\text{sq.\:deg}$. In the panel for $D_L$, we also show the fraction of detected sources within the threshold when the weak-lensing error is excluded.}
    \label{fig:err_rates}}
\end{figure*}

From each of the estimated posteriors, we derive the covariance matrix of the source parameters. The error in the measurement of a parameter $\theta$ can be described by $\Delta \theta=\sqrt{\text{var}(\theta)}$ where $\text{var}(\theta)$ is the variance of the parameter $\theta$ computed from the posterior samples. We are mainly interested in the masses, the spins, the luminosity distance, and the sky position of the sources. The error in the sky position ($\Delta\Omega$), which is
very important for performing multi-messenger astronomy~\cite{Tamanini2016}, is calculated as $\Delta\Omega = -2\pi\log(1-p)\sqrt{\left(\text{var}(\lambda)\text{var}(\sin\beta) - \text{cov}(\lambda,\sin\beta)^2 \right)}$ \cite{Cutler:1997ta}, where $\lambda$ and $\beta$ are the longitude and latitude of the source in the sky and $p$ is the credibility level
(which we fix to 90\%). For the luminosity distance, in addition to the statistical error from the posteriors, the weak-lensing error is expected to be non-negligible. As in  Ref.~\cite{Tamanini2016}, we estimate it following Ref.~\cite{2010PhRvD..81l4046H}, which gives
\begin{equation}
    \sigma_{\text{WL}}(z) = D_L \times 0.066 \left(\frac{1-(1+z)^{-0.25}}{0.25}\right)^{1.8}\,,
\end{equation}
and we add this error in quadrature to the statistical error. Errors on individual masses, spins and sky position are presented without the effect of weak-lensing. The effect of the latter, however, is expected to be small, as these quantities have little correlation with luminosity distance.

Our choice to summarize the posteriors by their covariance matrix is
potentially problematic due to multimodalities. The posterior distribution
of the sky position of the MBH binaries detected by LISA may indeed be multimodal \cite{Marsat:2020rtl, Mangiagli:2022niy}. Depending on the parameters of the system,
one may observe up to 8 modes in the sky. This is due to the LISA antenna pattern functions, but the motion of the LISA instrument during observations and the frequency-dependent features of the instrument response at high frequencies can help to break this degeneracy. Indeed, the mode containing the true source parameter (the ground truth) typically has the highest posterior probability. Fig.~\ref{fig:multi-modal} shows an example of the multi-modality commonly seen in sky location. In other cases, one may  observe only bi-modality in $\beta$, or the degeneracy may even be entirely absent with only a single peak in the posterior. In our analysis, we restrict ourselves to the primary peak alone (shown in red in Fig.~\ref{fig:multi-modal}): we compute the location of the degenerate modes \cite{Marsat:2020rtl}, divide the sky into quadrants so that each quadrant contains only one mode, and then we restrict the posterior to the quadrant with the true sky position and compute the associated sky position error. As a proxy for the effect of multimodalities, in cases where the true sky location does not lie in the quadrant with the maximum probability, we discard the source from further analysis and assume that it is simply not localized. 
This treatment is approximate, but multimodal posteriors are expected to represent a minority of the loud, nearby sources that are the best candidates for multimessenger observations~\cite{Mangiagli:2022niy}.

Next, to characterize the LISA performance, we compute the fraction of the sources for which the parameter estimation
 can be performed better than  the given thresholds. More precisely, we compute the fraction of detected sources for which the true sky position lies in the same quadrant as the peak of the posteriors \textit{and}:
the masses can be estimated to 1\% level or better; the spins can be estimated to within an absolute error of 0.01; the
luminosity distance can be estimated to 10\% or better; or the sky position can be estimated with an error of less than 10 squared degrees.
These thresholds are meant to be at $1\sigma$, except for the sky position (for which we set $p=0.9$).
The results of this analysis are shown in Fig.~\ref{fig:err_rates}. As can be seen,
the masses can be estimated well (i.e. to the 1\% threshold set above or better)  for $\sim 20-80\%$
of the detections. This fraction is  
 lower  for the spins (with the primary spin $\chi_1$ better determined
than the secondary one, $\chi_2$).
Similarly, the  luminosity distance
can only be estimated to 10\% error in $\sim 1-30\%$ of the detected events, partly as a result of weak lensing (Fig.~\ref{fig:err_rates}
also shows results without the lensing error). The sky position is well estimated for  $\sim 2-70\%$ of the detected sources, depending on the model. 

The difference between different models depends on the details of the
populations that they predict. For instance, the Q3-d (K+16) model
predicts more sources at low $z$, which explains the larger fraction of
events with well-estimated distance and sky location. HS models predict larger SNRs, which favor precise parameter estimation, but LS models
can also lead to good determination of the component masses and spins, thanks to the inspiral of these sources being in the LISA band for a long time.
We also stress that although the fractions are shown in Fig.~\ref{fig:err_rates}
may appear low, they need to be multiplied by the event rates shown in Table~\ref{tab:lisarate}, and can therefore result in a significant number of well-characterized sources, especially for the  HS models.

\section{Conclusions}
\label{sec:conclusions}

In this paper, we have re-assessed the prospects for detection and parameter estimation of MBH binaries with LISA, accounting for the constraints on semi-analytic models for the formation and evolution of MBHs from the recent detection of a GW background from PTA observatories. This extraordinary discovery  seems to suggest that large delays 
between galaxy and MBH mergers are disfavored and that MBHs merge efficiently after galaxy mergers. Moreover, accretion onto MBHs seems to be more efficient than previously thought, leading to a larger local MBH mass function at high masses. We, therefore, analyzed  models that are in good agreement with PTA data and produced synthetic catalogs of MBH binaries to characterize their emitted GW signals for LISA.

We find that LISA will detect at least a dozen MBH mergers in 4 years if we include  models that struggle to reproduce the quasar luminosity function at $z=6$. However, this number rises to $\gtrsim 100$ if we exclude 
such models. In fact, models that are in better agreement 
with the quasar luminosity function at $z=6$ can predict from several hundred to a few tens of thousands of detections (in some cases; cf.  Table~\ref{tab:lisarate}), potentially raising the questions of whether one will be able to resolve these signals singularly.
It should be noticed that similarly high detection numbers 
for LISA have been recently found by Ref.~\cite{Steinle:2023vxs}
on the basis of agnostic and astrophysically informed 
merger rate models calibrated to the PTA data. In fact, the
event rates found by Ref.~\cite{Steinle:2023vxs} are even higher than ours,
which can probably be ascribed to the differences in the population modeling
(ours is based on a full-fledged semi-analytic galaxy formation model, while theirs
is based on a phenomenological parametrization of the merger rate).
 
We  find that the models with the highest detection rate  (namely HS models) also have larger SNR, which translates into better accuracy when performing parameter estimation. The fraction of detected sources for each model, satisfying given thresholds on the parameter errors, is summarized in Fig.~\ref{fig:err_rates}.
An important caveat is that we performed Bayesian parameter estimation on the single events, without accounting for the presence of the other (overlapping) MBH binaries. The superposition of hundreds or thousands of MBH  signals would pose a serious challenge 
to LISA data analysis, and the recent detection of a stochastic GW background from PTA experiments is warning us that this may be a concrete possibility.

We stress that the fact that the PTA data seem to favor models with short or no delays is surprising, but
not to the point of being (overly) concerning. First, as explained in Sec.~\ref{sec:sam}, even the no-delay  models do include
at least the timescale related to the dynamical friction (including tidal effects) between dark matter halos. That may be
slightly overestimated (due to the uncertainties in the underlying models~\cite{Boylan-Kolchin2008,Taffoni2003}), and at least partially compensate for additional
``baryonic'' delays. More in general, the latter are computed in our semi-analytic model based on simple  formulae~\cite{Antonini2015,Bonetti2018b,2019MNRAS.486.4044B,tremmel} calibrated to
simulations (which are themselves affected by uncertainties due to resolution and subgrid physics), which could cause significant errors. Finally, the no-delay and short-delay 
models being preferred may point to gas being
more important than previously thought in MBH mergers, at least at high masses and low redshift. This possibility
may be checked with future PTA measurements, as gas interactions, besides providing very short delays, also tend to
flatten the PTA spectrum~\cite{Bonetti2018b}.
Eventually, since the dynamics leading to the delays between galaxy/halo mergers and MBH mergers may (and probably will) be
different at high redshift and low masses than it is in the local Universe for MBHs in the PTA band, the ultimate and most complete answer to
this puzzle will be provided by the LISA observations themselves.

\appendix

\section{Merger and detection rates for different models}
\label{allmergedet}
\begin{figure*}
    \centering
    \includegraphics[width=0.49\textwidth]{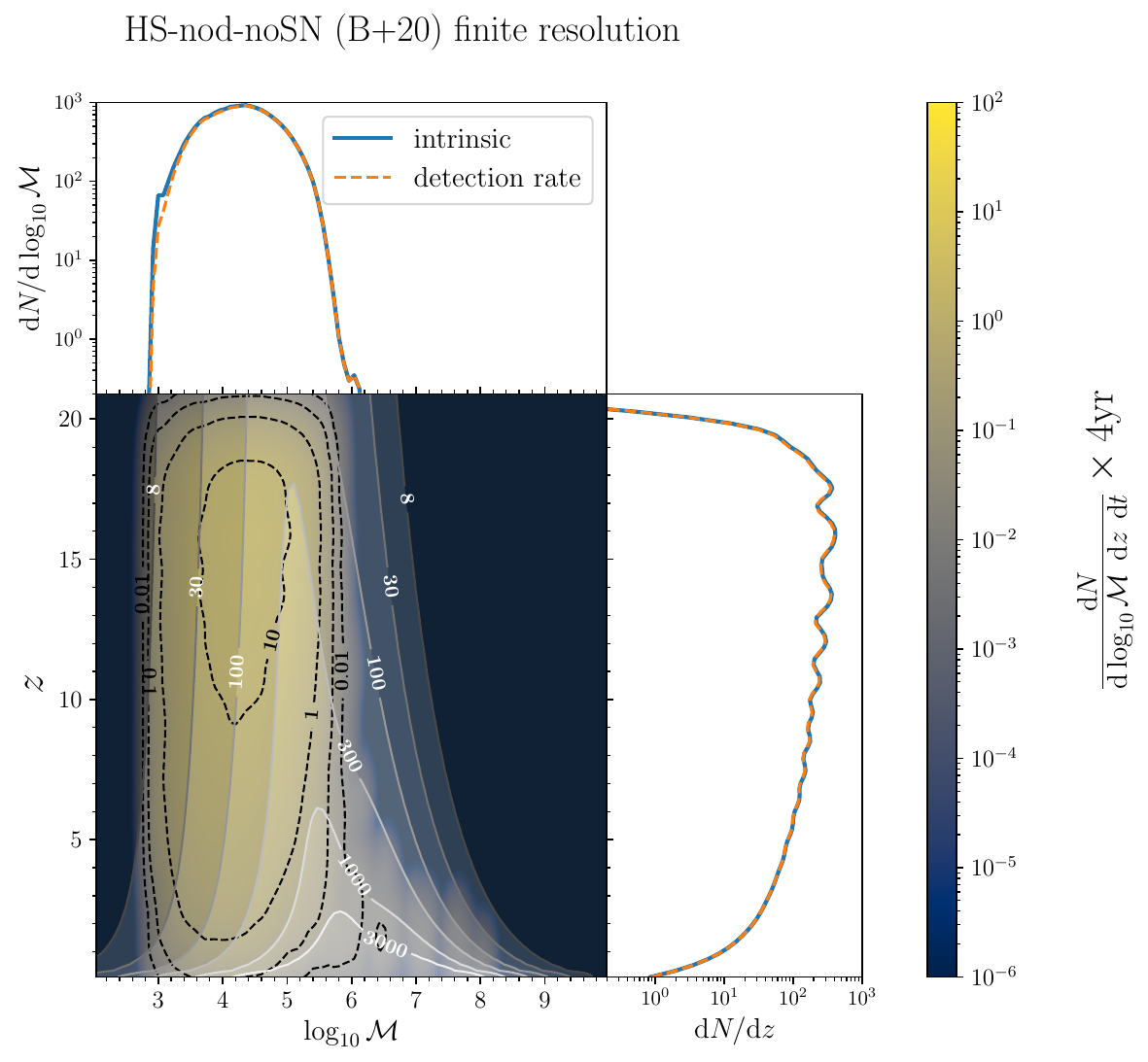}
    \includegraphics[width=0.49\textwidth]{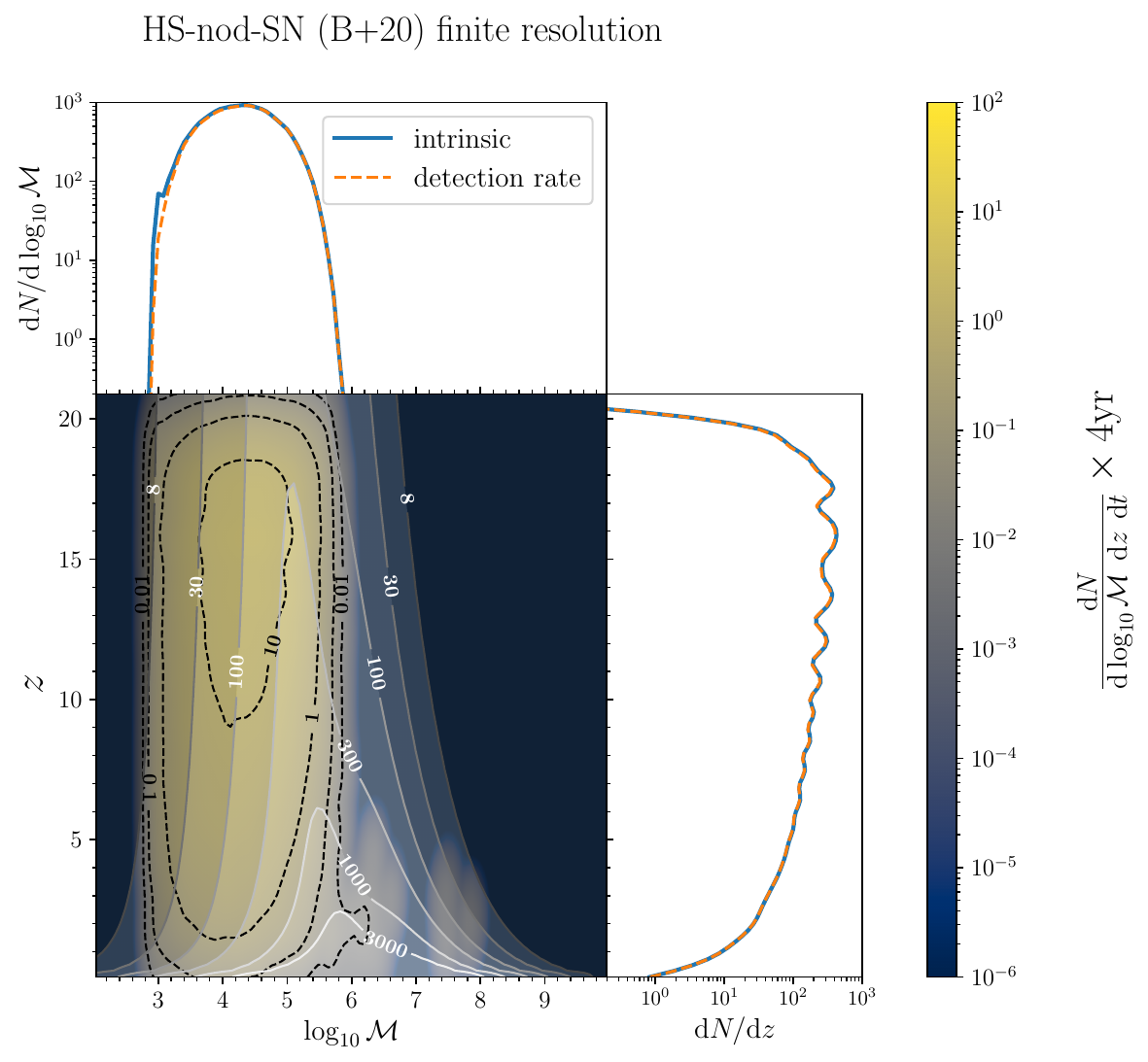}
    \includegraphics[width=0.49\textwidth]{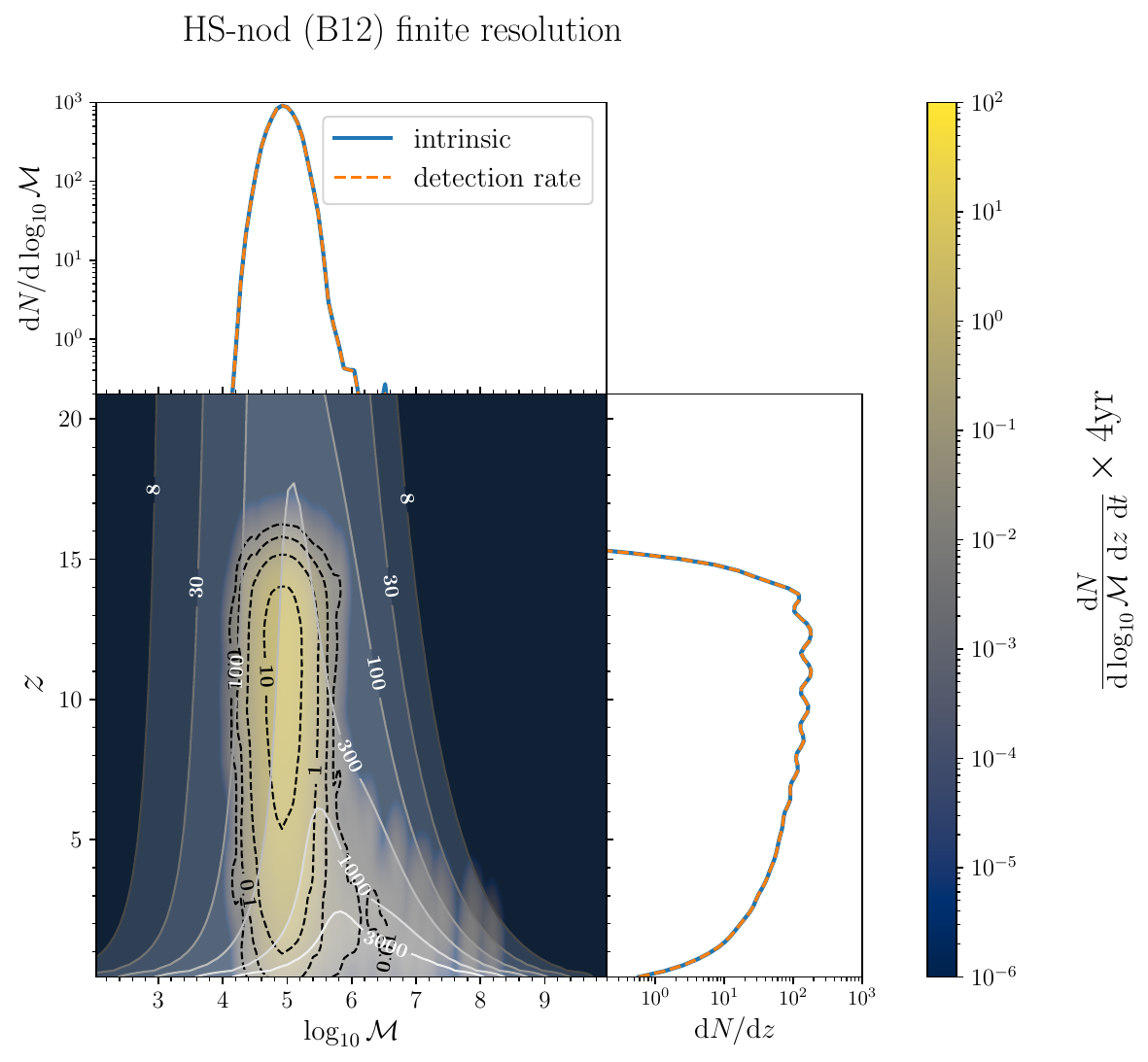}
    \includegraphics[width=0.49\textwidth]{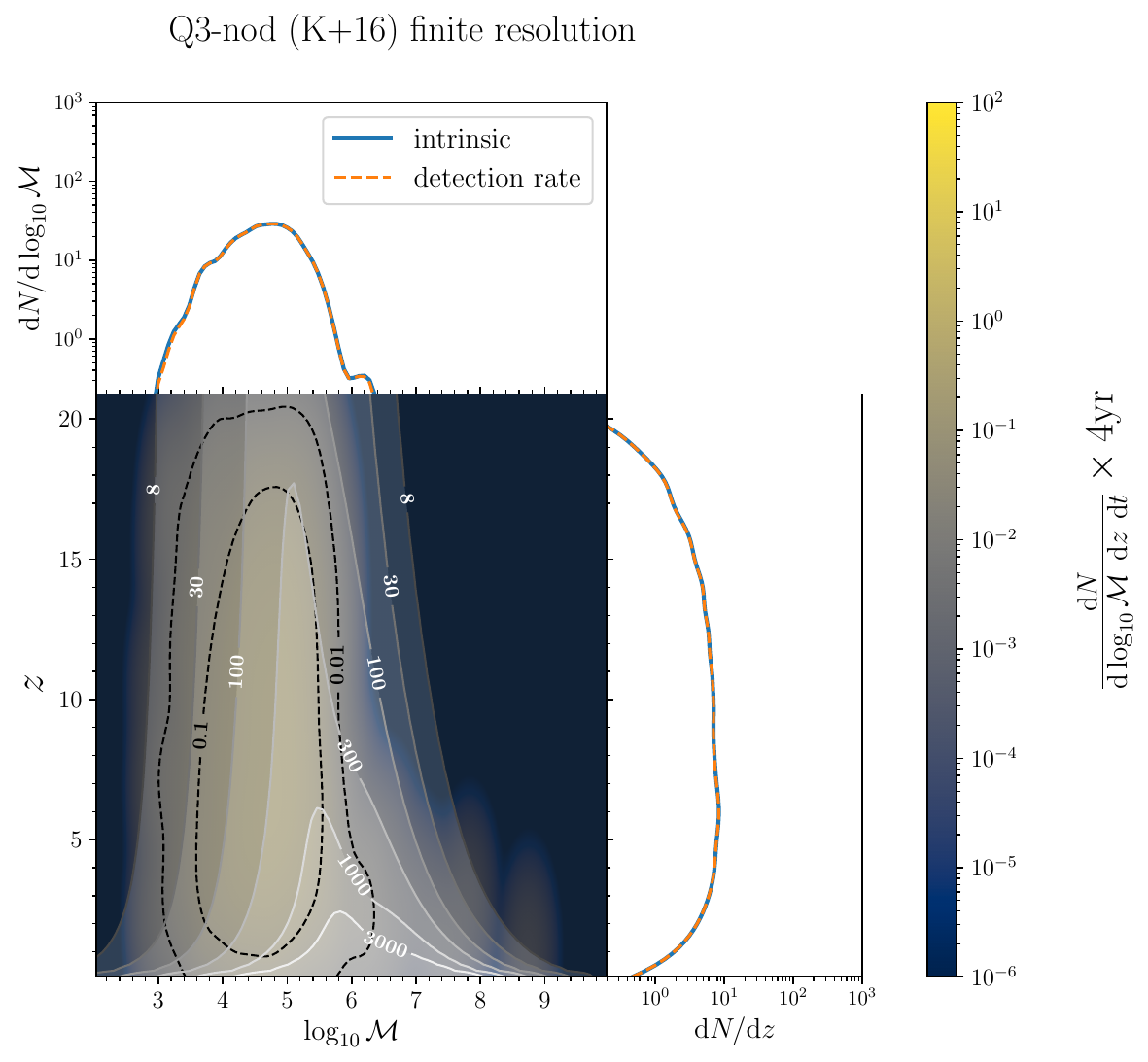}
    \includegraphics[width=0.49\textwidth]{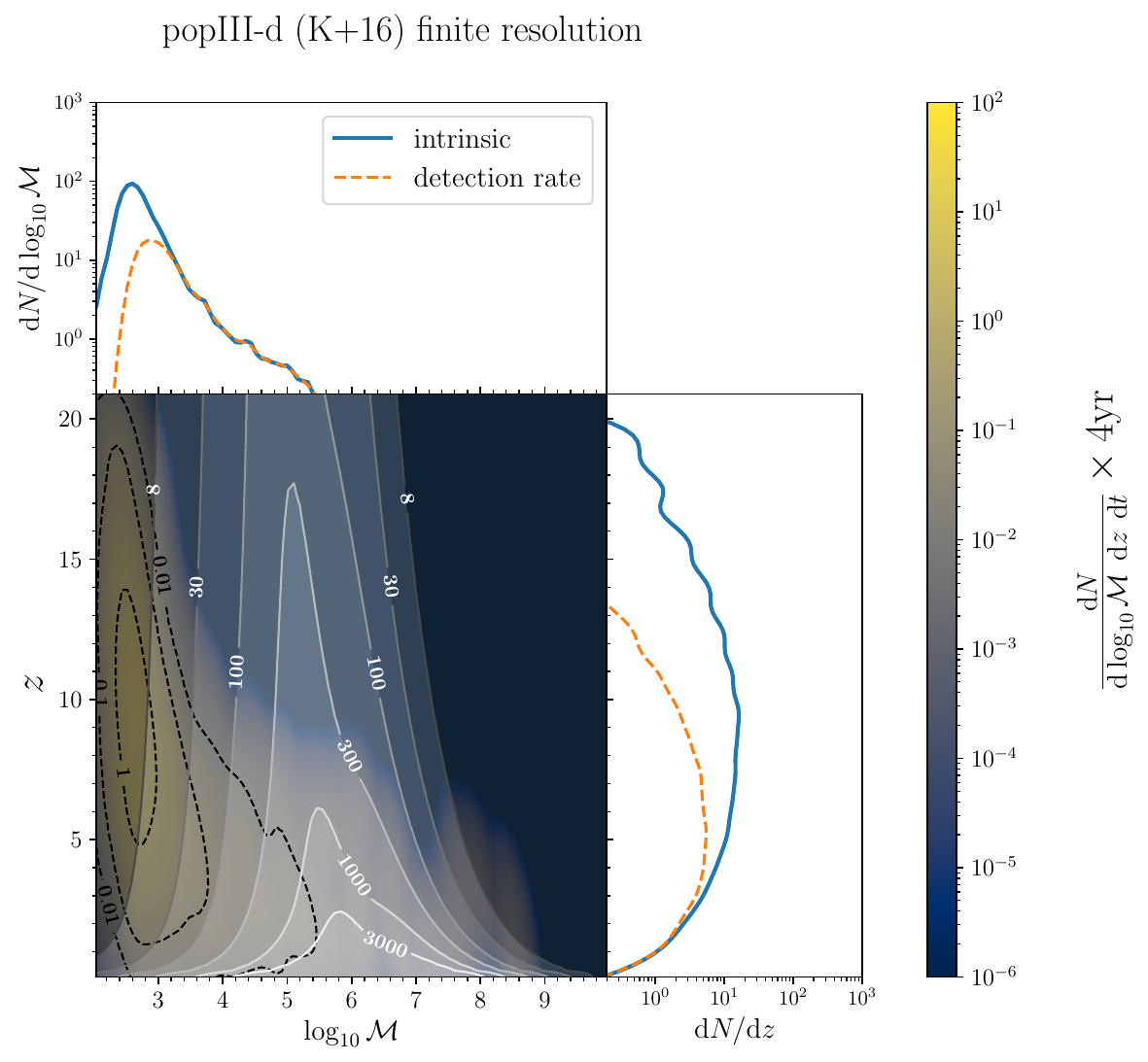}
    \includegraphics[width=0.49\textwidth]{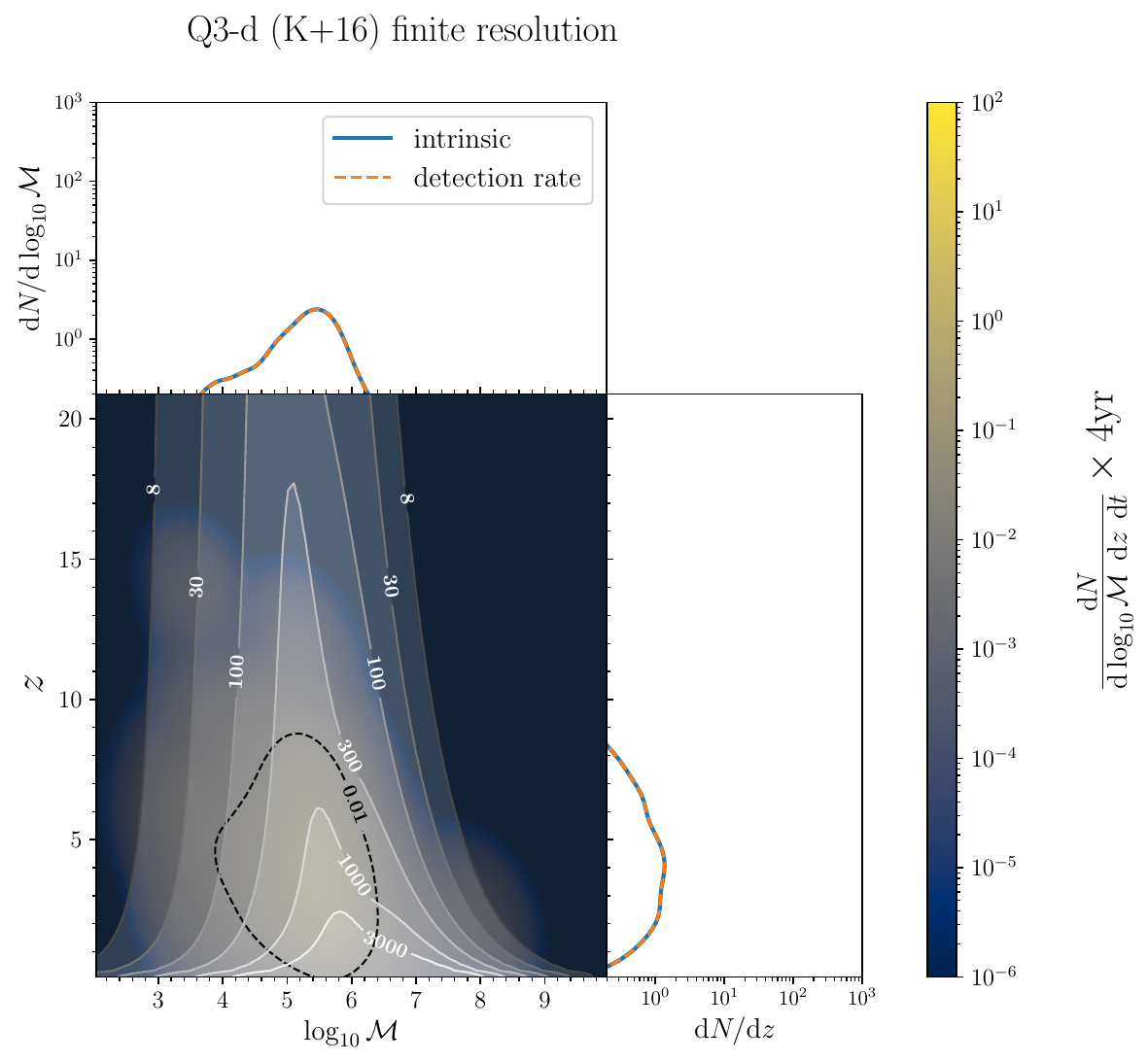}
    \caption{The same as Fig.~\ref{fig:rate1}, for the other models used in this work.\label{extrafig}}
\end{figure*}

In this Appendix, we show the mass and redshift distribution of
the total and the detected number of binaries, in Fig.~\ref{extrafig},
following the same format as in Fig.~\ref{fig:rate1}, but for the 
other models used in this work.

\acknowledgments
\no We thank A. Sesana for insightful conversations on the astrophysics of MBHs and PTA experiments. EB and MC acknowledge support from the European Union’s H2020 ERC Consolidator Grant “GRavity from Astrophysical to Microscopic Scales” (Grant No. GRAMS-815673) and the EU Horizon 2020 Research and Innovation Programme under the Marie Sklodowska-Curie Grant Agreement No. 101007855. KD, SB, and AP acknowledge IISER Thiruvananthapuram for providing high-performance computing resources at HPC Padmanabha.

\bibliographystyle{./utphys.bst}
\bibliography{main}

\end{document}